\documentclass[a4paper]{article}

\usepackage{hhline}
\usepackage{multirow}
\usepackage[margin=1.3in]{geometry}
\usepackage{lineno,hyperref}
\usepackage{natbib}
\usepackage{graphicx}
\begin{document}


\title{On-the-fly Detection of Autogenerated Tweets}
\author{Jonas Lundberg \\  \href{mailto:jonas.lundberg@lnu.se}{jonas.lundberg@lnu.se}
   \and Jonas Nordqvist
    \and Antonio Matosevic  }
\maketitle






%
%

\begin{abstract}
Most previous work related to tweet classification have focused on identifying a given tweet as a spam, or to classify a Twitter user account as a 
spammer or a bot. In most cases the tweet classification has taken place offline, on a pre-collected dataset of tweets.
In this paper we present an \emph{on-the-fly} approach to classify each newly downloaded tweet as \emph{autogenerated} or not. 
We define an autogenerated tweet (AGT) as a tweet where all or parts of the natural language content is generated automatically by a bot 
or other type of program.

Our on-the-fly approach makes use of two classifiers. The first classifies a tweet solely based on the twitter text and the tweet metadata 
that comes with every tweet. It is used for tweets posted by unknown users with no available tweet history. An unknown user also 
triggers a batch job to start downloading the missing user timeline information. The second classifier is used for tweets posted by a user 
where the user timeline is downloaded and available. Initially, it will be the first classifier that handles most of the tweets. This will gradually 
change and after an initialization phase where we download historic data for the most active users, we reach a state where the second classifier handles 
a vast majority of all the tweets.

A simulation using our on-the-fly detection mechanism indicates that we can handle Twitter streams with up to 68,000 unique users each day. The bottleneck is the time required to download new user timelines. The AGT detection is very accurate. In a set of 5,000 tweets we correctly classified about 98\% of all AGTs using a subject-wise cross-validation.
\end{abstract}



\section{Introduction}
\label{introduction}
The origin of this work is the cross-disciplinary Nordic Tweet Stream (NTS) project aiming at creating a multilingual text corpus consisting of tweets published in the five Nordic countries. 
The NTS linguists are explicitly interested in tweets having a text formulated by a human where each tweet is a personal statement. Tweets generated by bots and other programs or apps should be avoided since they might skew the results.

\citet{chuEtAl2012} classify Twitter user accounts as either a human, bot, or cyborg. The interesting part here is the cyborg category introduced to handle user accounts that contain both human and bot generated tweets. It could be an account mostly used by a bot posting weather forecast for Stockholm every hour where occasionally a human post a few tweets, or it could be an ordinary human user account where the user from time to time uses (for example) a beer app to generate a prefabricated tweet saying ``Drinking a X by @USER1 at @USER2 - URL''. 

In this paper we are interested in on-the-fly detection of non-human generated tweets. On-the-fly indicates that we are classifying tweets in real time as part of the Twitter download pipeline. 
A major obstacle here is that many suggested tweet detection techniques require access to historic data for
individual user accounts. For example, access to a user's tweeting history can reveal non-human tweeting behavior such as  periodic tweeting  or posting an extreme number of tweets per day. The Twitter REST API\footnote{dev.twitter.com/rest} provides access to historic data (the user timeline), but downloading the user timeline on-the-fly is much too slow to work in any realistic Twitter download scenario.

On-the-fly also mean that we classify individual tweets rather than user accounts. Non-human generated indicates that we,  in addition to tweets generated by bot accounts, also would like to identify tweets generated by any other type of app or program. We will refer to these non-human generated tweets as \emph{autogenerated tweets (AGTs)} rather than bot generated tweets to emphasize the difference.

Our contributions in this paper are the following.
\begin{enumerate}
\item We present a supervised machine learning based tweet classifier that classifies individual tweets rather than Twitter user accounts.
\item We present a two-classifier approach for on-the-fly detection of tweets that is suitable in cases when the classification would benefit from having access to online information that is time consuming to download.
\item We present a simulation of the on-the-fly mechanism to estimate its performance, to identify performance bottlenecks, and to identify its limitations. 
\item We make a thorough evaluation of the classifier using three different machine 
learning models based on 10,000 manually labeled tweets.
\end{enumerate}

The novel research contribution is that we present a general approach for how to 
classify individual tweets on-the-fly as a part of Twitter download stream that 
can be used in cases when the classifier needs access to online information that 
is time consuming to download. Our application domain is to identify autogenerated
tweets in the NTS data stream but the approach is general enough to 
be useful in many similar cases. For example, for on-the-fly detection of spam that 
requires online access to databases keeping track of identified spam URLs.

\section{Related Work}
\label{related_work}

Twitter is a microblogging mechanism allowing users to exchange short 140-character messages called tweets\footnote{www.twitter.com}. In the first quarter of 2017 Twitter was ranked as the 11th most popular website in the world by the Alexa ranking\footnote{www.alexa.com/topsites} with an estimated 328 million users\footnote{www.statista.com} posting 500 million tweets each day\footnote{internetlivestats.com/twitter-statistics}.

The main reason why Twitter has been tapped into in various scientific projects is (apart from its widespread use) that it comes with an open policy allowing third-party tools or users to retrieve at most a 1\% sample of all tweets. This service is called the Twitter Streaming API\footnote{dev.twitter.com/streaming/overview} and it enables programmers to connect to the Twitter server and download tweets in real time. Furthermore, the Twitter timeline mechanism (a REST API) provides access to the 3,200 most recent tweets posted by a given user.

The easy access and wide spread use of Twitter has made it a popular instrument to investigate a number of phenomena including the Arab Spring \citep{campbell2011egypt}, several political campaigns \citep{gayo2011limits,tumasjan2010predicting,howardKollanyi2016}, as a tool to for emergency communication \citep{sakaki2010earthquake,suttonEtAl2008}, and  to predict stock market prices \citep{bollen2011twitter}. Recently in linguistics, there have been various successful attempts to build both mono- \citep{scheffler2014} and multilingual text corpora of tweets \citep{laitinenEtAl2017}.

The growing popularity of Twitter has also made it an ideal target of spams and automated programs, known as bots. 
A bot is a program that posts tweets and estimates give that about 5-10\% of all users are bots\footnote{www.nbcnews.com/business/1-10-twitter-accounts-fake-say-researchers-2D11655362}, and that bots generates about 20-25\% of all tweets posted on Twitter\footnote{sysomos.com/inside-twitter/most-active-twitter-user-data/}. A bot can be harmless (e.g. a bot posting weather forecasts for Stockholm every hour) but also malicious (e.g. try to influence the outcome of an election campaign). By coordinating massive tweeting campaigns of particular hashtags, bots have been able to influence election campaigns \citep{howardKollanyi2016} and the very influential trending topics list \citep{ratkiewiczEtAl2011}.

In many research areas making use of Twitter as an instrument to study a certain phenomena, bots is an annoyance that might skew the result of the study. And consequently, similar to spam detection, there have been a rather large number of research papers related to bot detection \citep{chuEtAl2012,chavoshiEtAl2016,morstatterEtAl2016,zhangPaxson2011,DARPA2016}. Many of the ideas used for bot detection are similar to those used for spam detection. That is, machine learning based on account properties and/or tweet metadata. Bot detection also often make use of statistical features (e.g. entropy or $\chi^2$-test, \citep{chuEtAl2012,zhangPaxson2011}) to identify periodic patterns in their tweeting behavior. 

Many papers related to spam and bot detection focus on classifying \emph{user accounts} as a spammer/bot or not. The spam detection community also have a few papers presenting on-the-fly (run-time) spam detection approaches that classify individual tweets as spam or not \citep{millerEtAl2014,chenZhang2015, leeKim2013,wangZubiagaEtAl2015}. 
In order to classify tweets in real-time, as a part of a Twitter stream, these
on-the-fly papers avoid costly online database accesses and base their classification 
only on the tweet metadata that comes with each tweet. 

We are not aware of any paper dealing with on-the-fly bot detection. The major problem here is the need for historical (timeline) data to compute properties
like tweets per day, or to identify a periodic tweeting behavior (once every hour). Thus, most  of the bot detection papers take an offline approach where they first collect a tweet dataset, then collect various information about the user accounts, and then classifies the user accounts as a bot or not. 

In this paper we present a two-classifier approach for on-the-fly detection of autogenerated tweets in a Twitter stream. 
The machine learning approach we use in our two classifiers is similar to the bot detection approach presented by \citet{chuEtAl2012}. The major difference is that we, by using two classifiers, can apply it on-the-fly on individual tweets rather than offline in search of bot accounts.


\begin{figure*}[t]
\begin{small} 
\begin{tabular}{llllll}
\hline
\textbf{Rank} & \textbf{Denmark} & \textbf{Finland} & \textbf{Iceland} & \textbf{Norway} & \textbf{Sweden}\\
\hline
1 & English (46\%) & Finnish (55\%) & Icelandic (46\%) & English (37\%) & Swedish (52\%)\\
2 & Danish (30\%) & English (26\%) & English (36\%) & Norwegian (31\%) & English (29\%) \\
3 & Spanish (2\%) & Estonian (2\%) & Spanish (2\%) & Danish (5\%) & Spanish (1\%) \\
4 & Norwegian (2\%)  & Russian (2\%) & French (1\%) & Spanish (2\%) & Arabic (1\%) \\
5 & Swedish (2\%) & Swedish (1\%) & German (1\%) & Swedish (2\%) & Turkish (1\%)\\
\hline
\end{tabular}
\end{small}
\caption{\label{top_five} Top-5 languages used in tweets in the five Nordic countries.}
\end{figure*}

\section{Background}
\label{background}

The Nordic Tweet Stream (NTS) is a cross-disciplinary project which downloads Twitter messages from
the five Nordic countries\footnote{Denmark, Finland, Iceland, Norway, and Sweden}\citep{laitinenEtAl2017}. The primary goal in NTS is to create a new multilingual text corpus reflecting the written language used in social media in the Nordic countries. A secondary goal is to study the global expansion and diversification of English. The NTS linguists are explicitly interested in tweets having a text formulated by a human. Autogenerated tweets should be avoided since they might skew the results in linguistic research activities assuming that each tweet is a personal statement.

NTS uses the free Twitter Streaming API to collect tweets by specifying a \emph{geographical region} covering the five Nordic countries. This type of geobounded download make use of \emph{device location} metadata that comes with each downloaded tweet to make sure that each tweet is from a certain region.
 Twitter users who do not want to share their location are not included and it is difficult to determine how many Nordic tweets that are missed due to this geolocation requirement. Various estimates \citep{GrahamHG13, laitinenEtAl2017} indicate that only about 1-5\% of all posted tweets are geolocated. Hence, the NTS tweets (about 36,000 tweets/day) are geolocated to the Nordic countries but they only represent a small fraction of all tweets posted in this region.

A first set of NTS results was published in \citep{laitinenEtAl2017}. For example, Figure \ref{top_five} shows the five most frequently used languages in each of the five Nordic countries. Taken across all five countries, English is the main language, and its share is 32.3\%. This is a surprisingly high number considering English is not an official language in any of the Nordic countries. It might be that people are using English to increase their influence and to reach a larger audience, but it might also be that English is the preferred language for tweet generating bots and apps. 

This paper presents an approach to detect autogenerated tweets intended to be used in the NTS project to classify each newly downloaded tweet.
Initially we focus on classifying tweets identified as in English by the Twitter language recognition tool. Adding support for other languages in the Nordic countries is future work.

\begin{figure*}[t]
\begin{small}
\begin{center}
\begin{tabular}{ p{5cm}  p{5cm}  c }
\hline
\textbf{Example tweet} & \textbf{Comment} &\textbf{Class}\\
\hline
I was out walking 8.02 km with \#something \#somethingelse https://somewhere.com  & This tweet is generated by an app and by adding `I was out walking' it adds natural language to the tweet.& AGT\\
\hline
New year perfect photo frame!!\#something \#somethingelse @location https://somewhere.com & This tweet is also generated by an app but not considered an AGT since it does not add any natural language. The natural language was originally produced in the app by the user.& HGT\\
\hline
Wind 0.3 m/s W, Barometer 1016.0 hPa Rising slowly, Temperature 1.2 $^\circ$C, Rain today 2.7 mm, Humidity 99\% & This tweet is generated by a weather bot posting a new forecast every hour. & AGT\\
\hline 
\end{tabular}
\end{center}
\end{small}
\caption{\label{AGTfig}Examples of AGTs and non-AGTs.}
\end{figure*}

\subsection{Defining Autogenerated Tweets}
We define autogenerated tweets (AGT) as tweets where all or part of the natural language
content is generated automatically by a bot, application or other type of program. 
Also, by definition we do not automatically include tweets posted by an application, we only include those for which the application infers natural language to the tweet. 
For example, a bot (or app) that is retweeting a non-AGT is \emph{not} producing a new AGT since it is not adding any natural language.

Thus, AGTs in our definition come in two flavors. Tweets generated from pure bot accounts, such as weather bots, job bots, news bots, etc. The other kind are tweets generated by applications and programs in an otherwise human account. 
Figure \ref{AGTfig} presents examples of AGTs and non-AGTs according to our definition. The complement of all AGTs is henceforth called human generated tweets (HGT).


\begin{figure*} [t]
\footnotesize
\begin{verbatim}
boolean isAGT(Tweet tw) {
    UserProfile profile = ProfileGenerator.getProfile(tw)   // Is profile available?
    if (profile == null) {
         ProfileGenerator.fetchProfile(tw)           // Download profile (separate thread)
         return NoProfileClassifier.classify(tw)     // No profile classification
    }
    else {
        if (ProfileGenerator.isOld(profile))
            ProfileGenerator.fetchProfile(tw)         // Update profile  (separate thread)
        return ProfileClassifier.classify(tw,profile) // Profile based classification
    }
}
\end{verbatim}
\caption {Outline of On-the-fly AGT Detection System.} 
\label{online_system}
\end{figure*}
\section{On-the-fly Classification}
\label{overview}

The goal is to classify tweets on-the-fly. That is, the classifier should be a part of the Twitter download pipeline and be able to classify each tweet as they appear in the stream.  As pointed out earlier, the major problem with this approach is that more advanced AGT detection techniques, similar to bot detection,  require access to historic data for individual user accounts. 
The Twitter REST API provides access to historic data (the user timeline), but downloading the user timeline on-the-fly is much too slow to work in any realistic Twitter download scenario.

Figure \ref{online_system} shows schematic overview of our tweet classification system. The basic idea is to use two AGT classifiers: a no-profile based classifier (\texttt{NoProfileClassifier}) and  a profile based classifier (\texttt{ProfileClassifier}). A \texttt{UserProfile} is a set of properties related to a given user account that can be computed once the user timeline has been downloaded. These properties will be presented in Sections \ref{tweetProperties} and \ref{profileprop}, but in short: statistical measures to detect non-uniform (periodic) tweeting pattern as well as several ratios (e.g. URL, retweet, hashtag).

The algorithm in Figure \ref{online_system} starts by checking if such a profile is available. If not available ($\mathtt{profile == null}$) we first instruct the \texttt{ProfileGenerator} to start downloading the data required to compute a user profile. While downloading take place in a separate thread (\texttt{fetchProfile(tw)}), we handle the tweet in the best possible way without having access to user profile information. The \texttt{NoProfileClassifier} makes only use of the Twitter text and the metadata that comes with each individual tweet when classifying a tweet as AGT or not. 

The second part of the algorithm shows our handling of tweets for which a user profile is available. We first check if the profile starts to get ``old'', in which case we download  a new updated profile. Independent of the profile being old or not, we apply our second AGT classifier, the  \texttt{ProfileClassifier}, using the current user profile as well as the current tweet as input. 
The Profile Classifier has access to all the information available to the no-Profile Classifier as well as the user profile information.

The basic idea is rather simple. We use our best-effort Profile Classifier whenever possible, the no-Profile Classifier when no profile is available, and we download new user profiles as a batch job on demand.

\begin{figure*}
\hfill
\begin{minipage} {0.48\linewidth}
\includegraphics[scale=0.48]{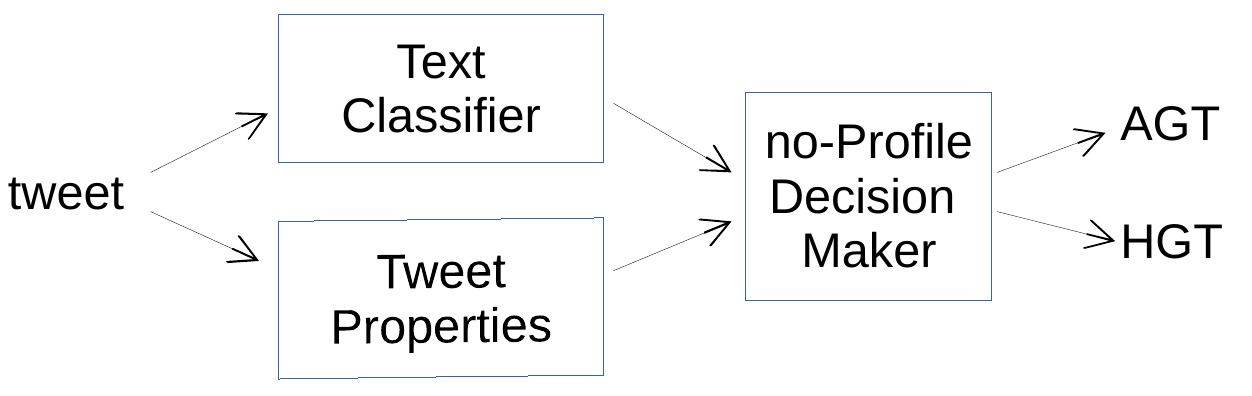}
\end{minipage}%
\hfill
\begin{minipage}{0.48\linewidth}
\includegraphics[scale=0.48]{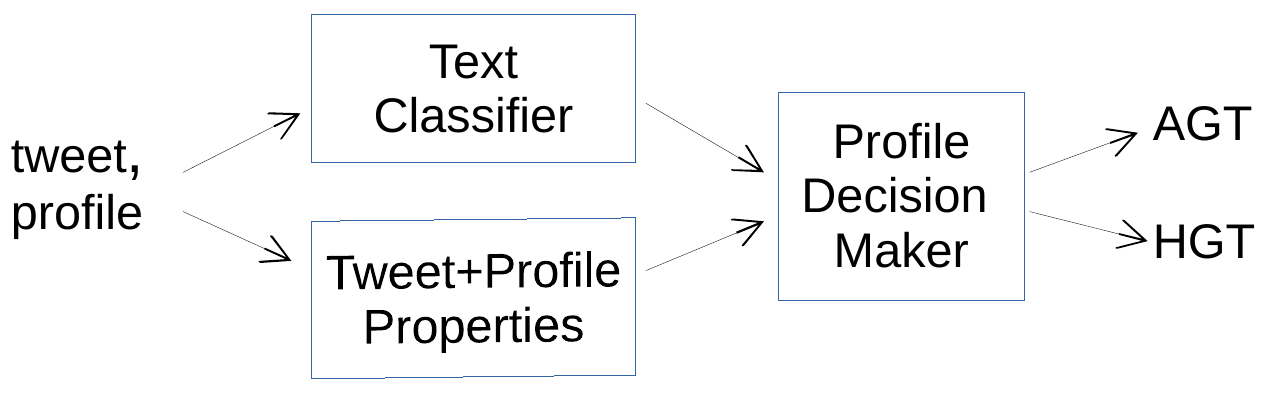}
\end{minipage}
\caption {Overview of the no-Profile Classifier (left) and the Profile Classifier (right). } 
\label{two_classifiers}
\end{figure*}

Figure \ref{two_classifiers} shows an overview of the no-Profile Classifier (left) and the Profile Classifier (right). They have a very similar structure. 
The actual tweet text is send to the \emph{Text Classifier} that uses machine learning to estimate the probability that a given tweet is autogenerated. The process is similar to machine learning based bot detection (e.g. \citep{chuEtAl2012}) although the Text Classifier is trained for another purpose. The machine learning approach used in the Text Classifier is described and motivated in Section \ref{TCsec}.

Each classifier also has a property component that computes a number of numerical properties for each tweet. The no-Profile property component computes nine properties based only on the metadata that comes with every tweet. The Profile property component computes the same nine properties and eleven additional properties based on user timeline information. The property components do not involve any machine learning, they simply compute a number of straight-forward numerical properties based on the available information.  The properties used in each property components will be presented in more detail in Sections \ref{tweetProperties} and \ref{profileprop}.

The final part in each classifier is the \emph{Decision Maker}. For each tweet the Decision Maker receives tweet text classification information from the Text Classifier, and a number of numerical properties from the property component.  The Decision Maker also makes use of machine learning and will be presented in more detail in  Sections \ref{noProfileClassifier} and \ref{profileClassifier}.

\section{User Profile Download}
\label{profile_download}

\begin{figure*}[t]
\includegraphics[scale=0.50]{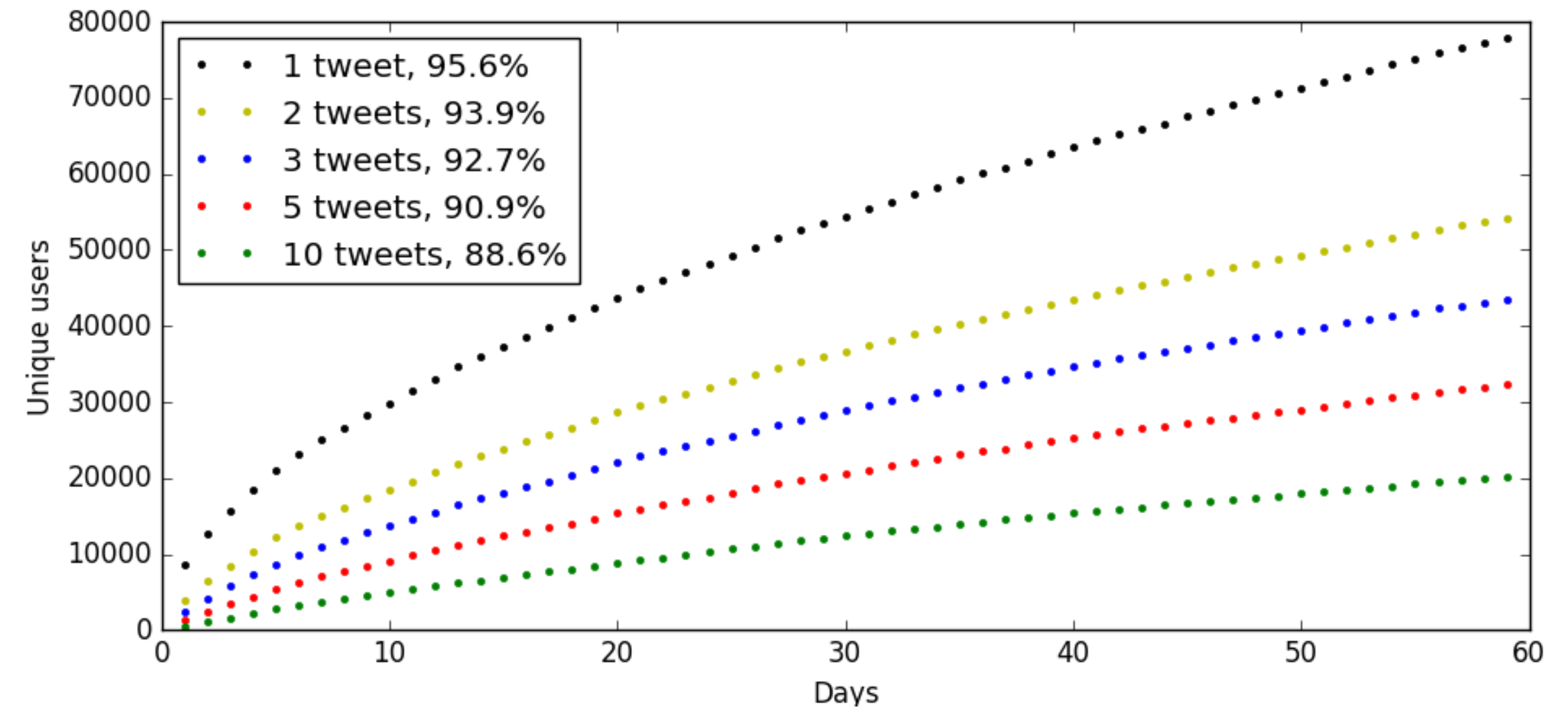}
\caption {Download requirement (per day) for a two months period.} 
\label{download_requirement}
\end{figure*}

A basic idea in our AGT detection approach is to download user profile information on demand when a new user is first discovered. 
Figure \ref{download_requirement} shows a simulation based on NTS data for a period of two months (Jan - Feb 2017). The simulation assumes that we started our AGT detection system on January 1 (Day 1) and that the profile download is instantaneous and take place directly after we have discovered a new user for the first time. That is, the first tweet is handled by the no-Profile Classifier and all the following tweets by the more technically advanced Profile Classifier. The plot named  \emph{1 tweet, 95.6\%} then shows the total number of new unique users for each day. The number of unique users discovered after Day 1 is 8,559 and the total number of unique users discovered after Day 59 is 77,845. The percentage 95.6\% gives the percentage of all tweets that was handled by the Profile Classifier (assuming that the no-Profile Classifier were used only once for each unique user).  As expected, the number of new unique users grows quickly in the beginning and flattens out after a while as our set of active users becomes more complete. But still, on Day 59 we discovered 617 new users that had not posted a single NTS tweet during the previous 58 days. 
The number of new users discovered Day 59, and the fact that the \emph{1 tweet} plot continues to increase steadily after two months, 
indicate that we have a rather large set of passive users that tweets rather seldom, maybe only once a month.

The passive users are most probably harmless when it comes to polluting the NTS dataset with autogenerated tweets. However, if taken into account they will force the user profile downloading mechanism  to  perform a costly download just to increase the precision when classifying their next tweet which might appear one month later (or never again). One way to reduce the number of downloads is to only download user information for user $X$ once we have seen $N>1$ tweets posted  by  $X$. The remaining plots \emph{N tweets, YY\%} in Figure \ref{download_requirement}  show the total number of new unique users that (since Day 1) have posted N or more NTS tweets. The percentages \emph{YY\%} show the percentage of all tweets that was classified using the Profile Classifier. For example, the plot \emph{3 tweets, 92.7\%} show that we after Day 1 (Day 59) had identified 2,340 (43,550) unique users that had posted at least three tweets, and that when the first three tweets are classified using the no-Profile Classifier,  92.7\% of all tweets posted during this entire two months period was handled by the more advanced Profile Classifier. Hence, using a threshold of 3 tweets will reduce the number of required downloads by 44\% at the cost of increasing the number of tweets handled by the less precise no-Profile Classifier by about 3\%.  

Two problems when downloading user profiles are 1) the actual time it takes to download the user timeline from the Twitter API server, and 2) the download restrictions that comes with the non-comercial version of the Twitter API. 
The timeline is a service that provides access to the 3,200 most recent tweets posted by the user.  Twitter uses a paging mechanism allowing a programmer to download a maximum of 200 tweets in each server requests, thus requiring about 16 server requests to fetch a single timeline. Furthermore, the non-commercial Twitter API imposes a rate limit  of a maximum of 900 server requests in each 15 minute window. After a few experiments we decided to use the following rather safe approach: We make 895 server requests and wait 5 minutes before we start to make another 895 server requests, etc. Using this approach we found that the average time to download the \emph{complete timeline} is about 15  seconds, or alternatively,  
about 5,770 new user profiles each day. 

Being able to download only 5,770 new user profiles each day would not be a major problem in the NTS case discussed earlier. If we start to download user timelines once we have seen 3 or more tweets from a certain user, we will already by the end of Day 1 have been able to download user information for all new unique users discovered that day. 

However, the NTS Twitter stream is due to its geolocation requirement very modest in size (about 36,000 tweets/day). In another Twitter related project we download all tweets having a text recognized as in Swedish by the Twitter language recognition tool. This stream averages about   205,000 tweets/day and have about  60,000 unique users each day\footnote{Also this stream is rather modest, it is quite possible to download millions of tweets per day and still be within the Twitter 1\% limit. }. The timeline download rate would in this case be a problem since it will take several days for the user information download mechanism just to handle all the new users discovered during Day 1, and consequently, that a large portion of all tweets should have to be handled by the simpler no-Profile Classifier.

A simple approach to handle the problem with the time consuming user timeline download is to limit the number of user timeline tweets we need to compute the user profile. If we decide to use only 200 tweets (one server request) rather than all tweets available, to compute a user profile, the cost for downloading reduces  significantly. Using the same approach (895 requests followed by a 5 minute break) we found that the average time to download the \emph{200 tweet timeline} is 1.27  seconds, or alternatively,  
about 68,000 new user profiles each day. Using 200 rather than 3,200 tweets to compute the user profile will of course in theory have a negative impact on the classification result. However, as we will see in Section \ref{profileClassifier}, the benefits of using all available tweets rather than a maximum of 200, is negligible. Hence, we will from now on base our user profile computations only on the 200 most recent tweets.
A stream with less than 68,000 new users each day will in that case eventually reach a steady-state where all new users are handled within one day of their discovery. We also use 200 tweets as our criteria for when identifying a user profile as ``old'' and in need to get updated (see Figure \ref{online_system}). 

Finally, the actual tweet classification is very fast compared to the user timeline download. A straight forward benchmarking of the three classifiers presented in Section \ref{overview}, where we repeatedly classified 10,000 tweets, give the following results.

\begin{small}
\begin{center}
\begin{tabular}{lr}
\hline
\textbf{Classifier} & \textbf{Classifications per second}\\
\hline
Text Classifier & 3,927\\
\hline
no-Profile Classifier & 2,830\\
\hline
Profile Classifier & 2,752\\
\hline
\end{tabular}
\end{center}
\end{small}

These results also include the time required to extract the wanted metadata from the tweets (represented as JSON objects), clean up the Twitter text (see Section \ref{TCsec}), and compute all the various properties used in the no-Profile and Profile property components.  When benchmarking the Profile Classifier we used pre-fetched user profiles. 

The ordering is of course the expected one since the no-Profile Classifier includes the Text Classifier, and the Profile Classifier includes all the properties in the no-Profile Classifier. Important to notice  is that the Profile Classifier performance (2,752 tweets/s) corresponds to about 238 million tweets per day. Hence, each of the these classifiers are fast enough to be a non-blocking component in a Twitter download pipeline using the non-commercial version of the Twitter Streaming API.

\section{Classification Components}
In this section we present the shared Text Classifier and the two property components mentioned in Section \ref{overview}. The Decision Maker components and the evaluation of the entire AGT detection system is presented in Section \ref{evaluation}.

\subsection{Text Classifier}
\label{TCsec}

Classifiers based on textual content have successfully been used in several studies  to detect spam and bots \citep{chuEtAl2012,wangZubiagaEtAl2015}. 
As our data contains several bots (e.g. weather bots, job bots, etc.) that repeatedly post tweet texts with a very similar content it is reasonable to believe that a classifier based on textual content could serve as a good indicator of AGTs.  Also, several of the AGTs posted by humans through applications share textual patterns that can be exploited by a text-based classifier. 

We apply traditional techniques for text-based classifiers to train a supervised machine learning model to recognize the AGTs in our data set. In Section \ref{evaluation} we will evaluate three such models. Independent of what model we choose, the textual content of a tweet is represented using a bag-of-words model based on the most frequent words in our data set. Besides the text content of the tweet, the input also includes predefined Twitter entities such as hashtags, user mentions, URLs and media (e.g. images, videos). All these entities are replaced by a token e.g. {\small{\tt{xhashtagx}}}, which are treated as regular words by the classifier. We also prepare the tweets prior to training by applying a stemming algorithm and removing all inter-punctuation in our tweets. It should also be noted that emojis are kept as is (i.e. each emoji is interpreted as a single word), as we believe that they might carry valuable information for the classifier. 

The result of these manipulations is that every tweet text is transformed into a set of words where each identified Twitter entity is replaced with a corresponding placeholder. For example, the Twitter text ``You will be greatly missed @POTUS !! https://t.co/BBpHaCvoV7''
 will generate the following set of words \{you, will, be, great, miss, xuserx, xurlx\}
 and this will be our input to the text classifier. The output of the Text Classifier, that will serve as input to the Decision Maker, is the estimated probability that a certain tweet is an AGT. Henceforth, this is called a tweet's \emph{text probability}.

\subsection{Tweet Properties}
\label{tweetProperties}

The Tweet Property component in the no-Profile Classifier computes nine  \emph{tweet properties}  attaining numerical and nominal values that can be computed directly from the tweet metadata. They will be used in the Decision Maker component together with the \emph{text probability}.
For instance, one should expect that humans have more followers than bots, or that AGTs tend to contain more URLs. 
Many of the properties used in the no-Profile Classifier, as well as in the Profile Classifier, have been discussed in \citep{chuEtAl2012,zhangPaxson2011,leeEoffCaverlee2011,DARPA2016}. The nine used properties are:

\begin{itemize}
\setlength\itemsep{0.1ex}
\item \textit{isReply}
	- binary variable indicating if the tweet is a reply to another tweet
\item \textit{isRetweet}
	- binary variable indicating if the tweet is a retweet
\item \textit{hashtagDensity, urlDensity, mentionDensity}
	- number of hashtag/URL/mention entities, respectively, in the tweet divided by the total number of the words in the tweet
\item \textit{accountReputation}
	-  number of followers divided by the number of friends and followers
\item \textit{tweetsPerDay}
	- total number of user's tweets divided by account age in days
\item \textit{favoritesPerDay}
	- number of tweets favorited by user divided by account age in days
\item \textit{deviceType}
	-  nominal variable based on the type of source used to post the tweet:
	\begin{enumerate}
	\setlength\itemsep{0.1ex}
		\item mobile: Twitter for Iphone, Twitter for Android etc.
		\item web: Twitter Web Client, Tweetbot for Mac etc.
		\item app: Instagram, Tumblr, Foursquared etc.
		\item SMM: Falcon Social Media Management, TweetDeck, dlvr.it etc.
		\item bot: Trendsmap Alerting, SpotifyNowPlaying, etc.
		\item other: newly observed not classified sources.
	\end{enumerate}
\end{itemize}
The tweet metadata contains as an attribute the source from which the tweet was posted, e.g., the name of an app, a program or a device. We manually classified the 140 most frequently used sources in our training set in one of the five categories, (1) $-$ (5), defined in the \emph{deviceType} attribute. These 140 sources cover about 95\% of all sources in the training set, while the remaining (unlabeled) sources were automatically classified as \emph{other}. The device type SMM stands for Social Media Management. That is, tools for managing content on multiple accounts on social networks.

\subsection{User Timeline Properties}
\label{profileprop}

Once the user timeline is downloaded we can define eleven additional properties that, together with the \emph{text probability} and the \emph{tweet properties}, are used in the Profile Classifier to detect AGTs. Note that the abbreviation \emph{RATs}, REST API Tweets, refers to the 200 most recent user tweets retrieved from the timeline part of the REST API.

\begin{itemize}
\setlength\itemsep{0.1ex}
\item \emph{hashtagRatio, urlRatio, mentionRatio}
	- the number of hashtag/URL/mention entities, respectively, in RATs divided by the number of RATs\
\item \emph{retweetRatio, replyRatio}
	- the number of retweets/reply tweets, respectively, in RATs divided by the number of RATs
\item \emph{mobileSourceRatio, webSourceRatio, appSourceRatio, smmSourceRatio, botSourceRatio, otherSourceRatio}
	- the number of tweets in RATs posted using source type  1, 2, 3, 4, 5 or 6, respectively, divided by the number of RATs (for source types see Subsection \ref{tweetProperties})
\item \emph{entropyHour, entropyMin, entropySec} - described below
\item \emph{chiSquaredMin, chiSquaredSec} - described below
\end{itemize}

Although many of the above properties are clearly correlated with the \emph{tweet properties}, we include them to resolve cases where more information about the user might be useful.

Two well-known statistical tools to detect regular tweeting behavior are so-called \emph{Pearson's $\chi^2$-$test$} \citep{zhangPaxson2011}  and \emph{entropy}  \citep{chuEtAl2012}. For the \emph{$\chi^2$-$test$} we assume that both minutes and seconds of tweets posted by humans follow a uniform distribution. If either of them fails to do so, we can suspect the user is not a human. This assumption however does not hold for hours since humans do not post uniformly throughout the day because they sleep, work or study. We use \emph{$\chi^2$-$test$} in order to test how well minutes and seconds of our tweets fit the uniform distribution defined as
\[\chi^2 = \sum \limits_{i = 1}^n \frac{(O_i - E_i)^2}{E_i},\] 
where $O_i$ is the observed frequency in the $i$-th bin and $E_i$ is the expected (theoretical) frequency in the $i$-th bin. Note that $E_i$ is the same for all bins $i$ since we expect a uniform distribution.
For each user, we calculate this number for both minutes and seconds and the corresponding $p$-values are used as attributes $chiSquaredMin$ and  $chiSquaredSec$. 

The second approach is \emph{entropy}, which can be thought of as a measure of complexity \citep{coverThomas2006}. The idea is to calculate time differences between every two consecutive tweets and compute the entropy for hours, minutes and seconds ($entropyHour$, $entropyMin$, $entropySec$) of those time differences. The entropy $H(X)$ for a random variable $X$, in our case the time difference, is defined as 
\[H(X)= - \sum \limits_{i = 1}^n P(x_i)\log_2(P(x_i)),\] 
where $P(x_i)$ is a relative frequency in the $i$-th bin. 
We can expect humans to have high entropy (complex/random behavior), and bots to have low entropy (simple/regular behavior) \citep{GianWang2007}.

It can be empirically seen that both \emph{entropy} and the \emph{$\chi^2$-$test$} depend on the choice of the bins. In case of too many bins small changes in the data considerably affect results, while the results are rather crude and may not indicate periodicity in the case of too few bins. After a few tests we settled to use fifteen bins for both methods.
Finally, the goal with both \emph{entropy} and \emph{$\chi^2$-$test$} is to detect periodic behavior. Their individual pros and cons will be discussed in Section \ref{profileClassifier}.

\section{AGT Detection Evaluation}
\label{evaluation}
The AGT detection system presented in this paper consists of three different classifiers that each can be used to classify a tweet as AGT or not. 
 All three classifiers are trained using three different supervised machine learning models from the machine learning toolkit WEKA \citep{Hall2009}. Rather than evaluating all applicable machine learning algorithms in WEKA, we decided to use the following three: 1) Na\"{i}ve Bayes (NB), 2) Support Vector Machine (SVM), and 3) Random Forest (RF). They have all been used to handle similar problems \citep{chuEtAl2012, mccordEtAl2011,leeEoffCaverlee2011} and they represent three different categories of machine learning models.

The NB model is a probabilistic classifier working under the na\"{i}ve assumption that the features are all independent. However, a study have shown that NB may work well even though the independence assumption is violated \citep{rish2001}. The classification is done by the following procedure: given a tweet $t$, and a class $C$, then $t$ will be assigned to class $C$ if the conditional probability $P(C| t)$ passes a certain threshold. See for instance \citep{John1995} for details. 

 The idea behind SVM is to find the hyperplane with maximal margin to separate the two classes AGT and HGT. To capture any non-linearity in our data we apply a Gaussian kernel. 
 More detailed information about SVMs can be found in \citep{Hastie2009}.

Random Forest is an ensemble method, which utilize decorrelated decision trees to produce a consensus model. It has shown to be effective and has good variance reduction. For a more detailed description see \citep{Breiman2001}.

The metrics we use to evaluate the results of the classifiers are precision and recall. Precision is the positive predictive value, i.e. the proportion of correctly classified instances among the total number of instances classified as AGTs. Recall or true positive rate is the proportion of correctly classified instances among the total number of AGTs in the ground truth set. For completion we also include the value of the $F$-measure, the harmonic mean of precision and recall (see e.g. \citep{Powers2011}). 

\subsection{Ground Truth}
From our NTS data set we took a random sample of 10,000 tweets from a 12 days period starting on January 1, 2017. Using the textual content of these and the user's previous tweets we could annotate them as belonging either to class AGT (2,251, 22.5\%) or class HGT (7,747, 77.5\%). Two tweets were removed from the ground truth as the user at the time of annotation (April 2017) was removed from Twitter and it was impossible to assign it to either class by simply looking at the textual content. 

Our ground truth set is further divided into two sets each containing roughly 5,000 tweets. The first part is used for training the Text Classifier and the second half is used for training the two Decision Makers. The splitting is done to avoid bias versus the Decision Maker classification.

\subsection{Subject-wise Cross-validation}

The traditional approach to test the performance of a classifier is to apply cross-validation on the training data. For $n$-fold cross-validation the training set is divided into $n$ disjoint sets. The classifier is then trained using $n-1$ of these sets and tested with some choice of metric against the $n$th set. This procedure is repeated until all $n$ sets are tested. In this way all instances are used for training on $n-1$ occurrences and tested for in 1 occurrence.

As some part of our data is collected and classified on tweet-level and not user-lever, we are, if not handled correctly, in risk of overestimating our results if we evaluate the Profile Classifier on tweet-level. The problem is that some of the parameters depend only on the user (see for instance \emph{entropy} in Section \ref{profileprop}) and thus implying that we would partly be training and testing the same instance. This would in turn imply that any overfitting of the model would become hard to detect. Some of the users in our set have a large proportion of the total amount of tweets and by random sampling of tweets in each fold, tweets from the same user would most likely end up in several folds, and thus be used for both training and testing.

A solution to the problem is discussed in \citep{Saeb059774} (in the setting of medicinal records and patients) and regards ensuring that the training set and test set in each fold of the cross-validation are disjoint with respect to users. Adopted to our training set it works as follows: From the 5,000 tweets we extract the unique user-ids. This set (consisting of roughly 1,750 users) is divided in 10 equally sized folds. Next we map all tweets back to the users. By this construction we have 10 folds of tweets which are disjoint with respect to users that can be used for cross-validation. Constructed in this way all instances are used for training and any overfitting of the model is far more likely to be visible in our testing procedure.

\subsection{Text Classifier}
\label{texteval}

The Text Classifier results in terms of precision, recall, $F$-measure and confusion matrix are illustrated in Figure \ref{evalTextClassifier}.

\renewcommand{\arraystretch}{1.1}
\begin{figure}[h]
\centering
\begin{minipage}[t]{.5\textwidth}
\centering
\begin{tabular}{|c||c|c|c|}
\hline
\textbf{}          & \textbf{NB} & \textbf{SVM} & \textbf{RF} \\ \hhline{|=||=|=|=|}
\textbf{Recall}    & 0.805            & 0.861    & 0.963              \\ \hline
\textbf{Precision} & 0.999            & 1.000    & 0.992              \\ \hline
\textbf{$F$-measure} & 0.892 & 0.969 & 0.977\\\hline
\end{tabular}
\vspace{2mm}
\label{noProfileResults} 
\end{minipage}%
\hfill
\begin{minipage}[t]{.5\textwidth}
\centering
\begin{tabular}{lccc}
                                             & \multicolumn{1}{l}{}              & \multicolumn{2}{c}{predicted}                                         \\ \cline{2-4} 
\multicolumn{1}{l|}{}                        & \multicolumn{1}{c|}{\textbf{}}    & \multicolumn{1}{c|}{\textbf{AGT}} & \multicolumn{1}{c|}{\textbf{HGT}} \\ \cline{2-4} 
\multicolumn{1}{l|}{\multirow{2}{*}{actual}} & \multicolumn{1}{c|}{\textbf{AGT}} 

& \multicolumn{1}{c|}{1,089}          
& \multicolumn{1}{c|}{42}         

 \\ \cline{2-4} 
\multicolumn{1}{l|}{}                        
& \multicolumn{1}{c|}{\textbf{HGT}} 

& \multicolumn{1}{c|}{9}          
& \multicolumn{1}{c|}{3,859}          

\\ \cline{2-4} 
\end{tabular}
\vspace{3mm}
\label{noProfileConfusion} 
\end{minipage}
\caption{Precision, Recall, and $F$-measure for the Text Classifier models (left) and the Confusion Matrix for the best model (Random Forest). }
\label{evalTextClassifier}
\end{figure}

All models performed very well and the misclassifications are few. The results of the Random Forset model (RF) stands out as the best in this setting, even though it performs slightly worse on precision. Thus, RF was our model of choice to use as input for the Decision Makers. RF correctly classified 1,089 of the 1,131 AGTs in the training set. A majority of the errors are false negatives.
The problematic cases are of different type, but they are in most cases related to humans posting AGTs using different apps. Three examples of such tweets are listed below.
\begin{small}
\begin{center}
\begin{tabular}{ p{8.5cm}  p{3.0cm}  }
\hline
{\bf{Tweet}} & {\bf{Text Probability}}\\
\hline
I just finished {\tt{xnumberx}} of doing circuit training with {\tt{xhashtagx}} {\tt{xhashtag}} {\tt{xurlx}} & 47 \%\\
\hline
I'm at {\tt{xuserx}} in Kuopio, Eastern Finland w/ {\tt{xuserx}} {\tt{xurlx}} & 39 \% \\
\hline
Discover hotels around somewhere in Norway from {\tt{xnumberx}} EUR per night: {\tt{xurlx}} & 7 \%\\
\hline 
\end{tabular}
\end{center}
\end{small}

It should be noted that the classifier does not reject these examples immediately, but they do not pass the threshold for AGT ($>$ 50\%). Seeing that we use the text probabilities as input for the Decision Maker, the output can give useful information even though it does not pass the threshold for the Text Classifier.

Still, these examples illustrates the limitation of using only a text classifier to solve the problem of finding AGTs, and two messages can be very similar in terms of textual content, but still come from different classes. Hence, there is a need to input more information into the Decision Maker.

\subsection{no-Profile Classifier}
\label{noProfileClassifier}
The attributes used in the no-Profile Classifier are the nine \emph{tweet properties} defined in Section \ref{tweetProperties} and the \emph{text probability} output from the Text Classifier. The results in terms of precision and recall are illustrated  in Figure \ref{evalNoProfile}.

\renewcommand{\arraystretch}{1.1}
\begin{figure}[h]
\centering
\begin{minipage}[t]{.5\textwidth}
\centering
\begin{tabular}{|c||c|c|c|}
\hline
\textbf{}          & \textbf{NB} & \textbf{SVM} & \textbf{RF} \\ \hhline{|=||=|=|=|}
\textbf{Recall}    & 0.991            & 0.995    & 0.996              \\ \hline
\textbf{Precision} & 0.979            &  0.978    & 0.979              \\ \hline
\textbf{$F$-measure} & 0.985 & 0.986 & 0.987\\\hline
\end{tabular}
\vspace{2mm}
\label{noProfileResults} 
\end{minipage}%
\hfill
\begin{minipage}[t]{.5\textwidth}
\centering
\begin{tabular}{lccc}
                                             & \multicolumn{1}{l}{}              & \multicolumn{2}{c}{predicted}                                         \\ \cline{2-4} 
\multicolumn{1}{l|}{}                        & \multicolumn{1}{c|}{\textbf{}}    & \multicolumn{1}{c|}{\textbf{AGT}} & \multicolumn{1}{c|}{\textbf{HGT}} \\ \cline{2-4} 
\multicolumn{1}{l|}{\multirow{2}{*}{actual}} & \multicolumn{1}{c|}{\textbf{AGT}} 

& \multicolumn{1}{c|}{1,097}          
& \multicolumn{1}{c|}{23}         

 \\ \cline{2-4} 
\multicolumn{1}{l|}{}                        
& \multicolumn{1}{c|}{\textbf{HGT}} 

& \multicolumn{1}{c|}{4}          
& \multicolumn{1}{c|}{3,875}          

\\ \cline{2-4} 
\end{tabular}
\vspace{3mm}
\label{noProfileConfusion} 
\end{minipage}
\caption{Precision, Recall, and $F$-measure for the no-Profile Classifier models (left) and the Confusion Matrix for the best model (Random Forest). }
\label{evalNoProfile}
\end{figure}

All models perform rather well with fewer misclassifications than the Text Classifier. Thus, a noteworthy improvement is obtained by including the tweet metadata. Since the Random Forest classifier is again, although marginally, the best out of the three tested models, we choose it for the result analysis. Three problematic cases are:
\begin{small}
\begin{center}
\begin{tabular}{p{8cm} l l }
\hline
\textbf{Tweet} & \textbf{Predicted} & \textbf{Actual}\\
\hline
Drinking a Fortunate Islands by {\tt{xuserx}} at {\tt{xuserx}} -- {\tt{xurlx}} & HGT & AGT\\
\hline
Listen to T. Cotugno - L'Italiano (Andre Rieu) HD by Abdulmajeed Zaro {\tt{xhashtagx}} on {\tt{xhashtagx}} {\tt{xurlx}} & HGT & AGT\\
\hline
Waiting for a plane (@ The Famous Bar in Stockholm-Arlanda, Stockholm) {\tt{xurlx}} & HGT & AGT\\
\hline
\end{tabular}
\end{center}
\end{small}

These three tweets were manually labeled as AGTs, but they were classified as HGTs. The reason being, besides low \emph{text probability}, the fact that they have been posted from rarely used apps which the no-Profile classifier has not been trained on. Moreover, the users seem to be humans who occasionally post such AGTs. Thus, neither attributes like \textit{accountReputation} or \textit{favoritesPerDay} can solve this problem. This is an issue from which the Profile Classifier suffers as well since introducing new features concerning the user does not capture these AGTs either.

\subsection{Profile Classifier}
\label{profileClassifier}
The attributes used in the Profile Classifier are the \emph{text probability}, the nine \emph{tweet properties} used in the no-Profile Classifier, and the eleven user timeline properties presented in Subsection \ref{profileprop}. The results obtained for the Profile Classifier are illustrated in Figure \ref{evalProfile}.

\renewcommand{\arraystretch}{1.1}
\begin{figure}[h]
\centering
\begin{minipage}[t]{.5\textwidth}
\centering
\begin{tabular}{|c||c|c|c|}
\hline
\textbf{}          & \textbf{NB} & \textbf{SVM} & \textbf{RF} \\ \hhline{|=||=|=|=|}
\textbf{Recall}    & 0.933            & 0.995    & 0.996              \\ \hline
\textbf{Precision} & 0.980            &  0.978    & 0.980              \\ \hline
\textbf{$F$-measure} & 0.956		 & 0.986 	& 0.988\\\hline
\end{tabular}
\vspace{2mm}
\label{ProfileResults} 
\end{minipage}%
\hfill
\begin{minipage}[t]{.5\textwidth}
\centering
\begin{tabular}{lccc}
                                             & \multicolumn{1}{l}{}              & \multicolumn{2}{c}{predicted}                                         \\ \cline{2-4} 
\multicolumn{1}{l|}{}                        & \multicolumn{1}{c|}{\textbf{}}    & \multicolumn{1}{c|}{\textbf{AGT}} & \multicolumn{1}{c|}{\textbf{HGT}} \\ \cline{2-4} 
\multicolumn{1}{l|}{\multirow{2}{*}{actual}} & \multicolumn{1}{c|}{\textbf{AGT}} 

& \multicolumn{1}{c|}{1,098}          
& \multicolumn{1}{c|}{22}         

 \\ \cline{2-4} 
\multicolumn{1}{l|}{}                        
& \multicolumn{1}{c|}{\textbf{HGT}} 

& \multicolumn{1}{c|}{4}          
& \multicolumn{1}{c|}{3,875}          

\\ \cline{2-4} 
\end{tabular}
\vspace{3mm}
\label{ProfileConfusion} 
\end{minipage}
\caption{Precision, Recall, and F-score for the Profile Classifier models (left) and the Confusion Matrix for the best model (Random Forest). }
\label{evalProfile}
\end{figure}

The results for the Profile Classifier do not differ much from the no-Profile Classifier when using SVM and RF models. Surprisingly, there is decrease in the performance of Na\"{i}ve Bayes, perhaps due to our increased violation of the independence assumption \citep{Zhang2004}. We analyze results for the best model, which is once again Random Forest. 

Since the no-Profile and Profile Classifier make 27 and 26 errors, respectively, one can say there is no great improvement by introducing eleven additional user properties. Moreover, most misclassifications in both approaches are the same tweets, nearly all of which are generated by rarely used apps and posted by humans. In addition, they have a low \emph{text probability.}

The main purpose of introducing user timeline data was to identify periodic tweets using the $\chi^2$-$test$ and $entropy$. The fact that the no-Profile and Profile Classifiers perform equally well does not imply that these statistical components do not work. As we have previously seen in \citep{zhangPaxson2011} and \citep{chuEtAl2012}, they do complete the task successfully. However, in our setting, it seems that most of such tweets have already been discovered using the Text Classifier since they often have similar text structures. For example, most of the frequently occurring weather bots use the same vocabulary and are easily detected by the Text Classifier.

As mentioned earlier, difficult cases are AGTs with low \emph{text probability} posted by humans from rarely used apps. We now consider two such tweets with similar properties, one of which is correctly classified, while the other one is not. 

\begin{small}
\begin{center}
\begin{tabular}{p{8cm} l l }
\hline
\textbf{Tweet} & \textbf{Predicted} & \textbf{Actual}\\
\hline
Rock Carvings in Tanum  {\tt{xhashtagx}}  {\tt{xurlx}} - visit now! {\tt{xhashtagx}} {\tt{xhashtagx}}. & HGT & AGT\\
\hline
Hammerfest Airport, Norway - Woodward Field, United States HFT-CDN Weather (Sat): snow/$-2^{\circ}$C.& AGT & AGT\\
\hline
\end{tabular}
\end{center}
\end{small}

Both tweets are AGTs generated by periodic bots $-$ the former tweet by a bot that posts information about tourist attractions every eight hours sharp and the latter tweet by a bot the posts weather reports from different locations every two hours sharp.
Both AGTs were assigned low \emph{text probability}, 0.06 and 0.07 respectively, characteristic to HGTs. On the other hand, $chiSquaredMin$ and $chiSquaredSec$ being almost 0 and low \emph{entropy} related attributes clearly indicate a periodic behavior typical for AGTs. The \emph{deviceType} attribute is $other$ according to Section \ref{tweetProperties} for both tweets, which does not help in the classification. Even though similar, the former tweet was misclassified with probability to be AGT of 0.48, while the latter one was correctly classified with probability of 0.58.  In such border cases other attributes contribute to one class prevailing over the other and more information about user can be helpful.

We included both $\chi^2$-$test$ and $entropy$ as statistical tools to identify a periodic tweeting behavior in the Profile Classifier. In our setting both of them complete the task well. When one method works, the other does as well. 
However, in a more general setting it might be better to use $entropy$ as a measure of irregularity since it does not assume any underlying distribution, while $\chi^2$-$test$ does.

We tested the Profile Classifier using features computed with the help of both 200 and 3,200 of the user's most recent tweets. Since the difference between results was negligible, we decided to use the approach with the 200 most recent tweets in order to avoid a bottleneck when downloading timelines.

\subsection{Attribute Selection}

Similar results for the no-Profile and Profile Classifier lead us to assume that some common attributes contribute significantly to the result. In order to identify them we apply the Attribute Selector in WEKA \citep{Hall2009} in each fold to avoid bias. The chosen attribute evaluator was CfsSubsetEval, which prefers attributes having a high correlation with the class itself and low intercorrelation. 
Four attributes are found to be important for both no-Profile and Profile Classifier, namely \textit{textProbability},  \textit{deviceType},  \textit{accountReputation} and  \textit{favoritesPerDay}. We also find \emph{retweetRatio} to be important for the Profile Classifier.

We have already seen in Section \ref{texteval} that the  \textit{text probability} itself gives a rather good results. Also, the importance of  \textit{deviceType} attribute means that the HGTs and AGTs are mostly posted from different sources. Attributes \textit{accountReputation} and \textit{favoritesPerDay} emphasize well the difference between classes since they are both expected to be high for HGTs and low for AGTs. The additional attribute \textit{retweetRatio} reveals that there are many AGTs that mostly retweet in order to spread some idea or advertisement \citep{Varol2017}. 


\section{Summary, Conclusions, and Future Work}
\label{summary}
In this paper we present a two-classifier approach for on-the-fly classification of tweets that is suitable in cases when the classification benefits from having access to online information that is time consuming to download. Our application domain is to identify autogenerated tweets (AGT, non-human generated) in the NTS data stream (see Section \ref{background}) and  historic user account data (the \emph{user timeline} from the Twitter REST API) is the online data we need to download in order to make the best possible classification. 

The two-classifier approach (see Section \ref{overview}) is rather straight forward. We have a ``simple'' classifier (the \emph{no-Profile Classifier}) that can classify a tweet only based on the metadata that comes with every tweet, and a more ``advanced'' classifier (the \emph{Profile Classifier}) that makes use of downloaded user timeline information. The no-Profile Classifier is used for ``new'' users where no user timeline is available. The discovery of a new user also triggers a batch job to download the missing user timeline. The more advanced Profile Classifier is used for each tweet posted by a user where the user timeline is already downloaded. 

Section \ref{profile_download} presents a simulation of the two-classifier approach applied to two months (59 days) of NTS data. 
In summary, the two-classifier approach can handle Twitter streams with less than 68,000 unique users each day, and more than 93\% of all tweets in this scenario will be classified by the more advanced Profile Classifier. The bottleneck is (as expected) the time it takes to download new user timelines.

The two classifiers have a similar structure (see Section \ref{overview}). They both contain a Text Classifier, a property computing component, and a Decision Maker that makes the final classification (AGT or not AGT) based on the input provided by the Text Classifier and the property component.

Section \ref{evaluation} presents an evaluation of the AGT classifier using a manually labeled training set containing 10,000 tweets 
where 2,251 tweets (22.5\%) are labeled as AGTs. The evaluation also includes the Text Classifier since it can be used as a stand-alone tool for AGT detection. The share of correctly classified AGTs for the three classifiers are:
\vspace{\baselineskip} \noindent
\begin{center}
Text Classifier: 96.3\%, no-Profile Classifier: 97.9\%, Profile Classifier: 98.0\%.
\end{center}
\vspace{\baselineskip}
All AGT detectors (also the Text Classifier) performs very well. The difference between the Text Classifier and the no-Profile Classifier is noteworthy (24 less misclassifications) and mostly due to the \emph{deviceType} attribute which turns out to be a reliable indicator for identifying certain types of AGT generators that are hard to spot using only the textual content of the tweet. 

The difference between the no-Profile Classifier and the Profile Classifier is (somewhat surprisingly) negligible. The strength of the Profile Classifier, and the major reason for downloading the user timeline, is to identify periodic tweets using the $\chi^2$-$test$ and entropy. However, as it turns of in this particular NTS dataset, the Text Classifier is almost as good in detecting periodic tweets. The periodic tweets are numerous, often have a similar text structure, and as a result, the Text Classifier learns to recognize most of them during the training phase. 

One possible conclusion in this scenario (NTS data, AGT detection) is that one can skip downloading the user timeline and trust the no-Profile Classifier to identify periodic tweets. 
This would definitely simplify the handling and remove many of the performance bottleneck problems discussed earlier. However, it is reasonable to expect the performance of the Text Classifier and no-Profile Classifier to deteriorate over time as new bots, and new apps and programs for tweet publication, show up.  In this scenario the Profile Classifier that downloads new user timeline information regularly is more likely to identify (in particular) periodic bots. Measuring this expected deterioration and applying techniques of continuous learning to avoid it is future work.

A closer look at the tweets that are misclassified by both classifiers show that many of them are similar. They are app generated tweets (from rarely used apps) posted by a human user, and with a text assigned a very low AGT probability by the Text Classifier. Although other attributes (e.g. \emph{account reputation}, \emph{device type}, or \emph{entropy}) might indicate an AGT, the classifier relies largely on the Text Classifier and incorrectly classify them as human generated.
Fine-tuning the impact of the Text Classifier is also future work.


\bibliography{twitter}

\end{document}